%% file: 0-main.tex
\pdfoutput=1
\documentclass[superscriptaddress, twocolumn, prd]{revtex4-1}

\usepackage{graphicx}
\usepackage{float}
\usepackage{caption}
\usepackage{subcaption}
\usepackage{changes}
\usepackage{multirow}
\usepackage{natbib}
\usepackage{hyperref}
\usepackage{amsmath}
\usepackage{color,colortbl}
\newcommand{\specialcell}[2][c]{%
	\begin{tabular}[#1]{@{}c@{}}#2\end{tabular}}
\begin{document}
\definecolor{lightgray}{gray}{0.9}
\author{Michael P. Ross}
\affiliation{Center for Experimental Nuclear Physics and Astrophysics, University of Washington, Seattle, Washington, 98195, USA}
\author{Timesh Mistry}
\affiliation{The University of Sheffield, Sheffield S10 2TN, UK}
\author{Laurence Datrier}
\affiliation{SUPA, University of Glasgow, Glasgow G12 8QQ, UK}
\author{Jeff Kissel}
\affiliation{LIGO Hanford Observatory, Richland, WA 99352, USA}
\author{Krishna Venkateswara}
\affiliation{Center for Experimental Nuclear Physics and Astrophysics, University of Washington, Seattle, Washington, 98195, USA}
\author{Colin Weller}
\affiliation{Center for Experimental Nuclear Physics and Astrophysics, University of Washington, Seattle, Washington, 98195, USA}
\author{Kavic Kumar}
\affiliation{University of Maryland, College Park, MD 20742, USA}
\author{Charlie Hagedorn}
\affiliation{Center for Experimental Nuclear Physics and Astrophysics, University of Washington, Seattle, Washington, 98195, USA}
\author{Eric Adelberger}
\affiliation{Center for Experimental Nuclear Physics and Astrophysics, University of Washington, Seattle, Washington, 98195, USA}
\author{John Lee}
\affiliation{Center for Experimental Nuclear Physics and Astrophysics, University of Washington, Seattle, Washington, 98195, USA}
\author{Erik Shaw}
\affiliation{Center for Experimental Nuclear Physics and Astrophysics, University of Washington, Seattle, Washington, 98195, USA}
\author{Patrick Thomas}
\affiliation{LIGO Hanford Observatory, Richland, WA 99352, USA}
\author{David Barker}
\affiliation{LIGO Hanford Observatory, Richland, WA 99352, USA}
\author{Filiberto Clara}
\affiliation{LIGO Hanford Observatory, Richland, WA 99352, USA}
\author{Bubba Gateley}
\affiliation{LIGO Hanford Observatory, Richland, WA 99352, USA}
\author{Tyler M. Guidry}
\affiliation{LIGO Hanford Observatory, Richland, WA 99352, USA}
\author{Ed Daw}
\affiliation{The University of Sheffield, Sheffield S10 2TN, UK}
\author{Martin Hendry}
\affiliation{SUPA, University of Glasgow, Glasgow G12 8QQ, UK}
\author{Jens Gundlach}
\affiliation{Center for Experimental Nuclear Physics and Astrophysics, University of Washington, Seattle, Washington, 98195, USA}

\title{Initial Results from the LIGO Newtonian Calibrator}


	\date{\today}

	\begin{abstract}

	The precise calibration of the strain readout of the LIGO gravitational wave observatories is paramount to the accurate interpretation of gravitational wave events. 
	This calibration is traditionally done by imparting a known force on the test masses of the observatory via radiation pressure. 
	Here we describe the implementation of an alternative calibration scheme: the Newtonian Calibrator. 
	This system uses a rotor consisting of both quadrupole and hexapole mass distributions to apply a time-varying gravitational force on one of the observatory's test masses. 
	The force produced by this rotor can be predicted to $<1\%$ relative uncertainty and is well-resolved in the readout of the observatory. 
	This system currently acts as a cross-check of the existing absolute calibration system.
	\end{abstract}
	\maketitle

	\input{2-Introduction}
	\input{3-Background}

\input{4-Hardware}

\input{6-Modelling}

\input{7-Measurements}
	\input{8-Results}
	\input{9-ConclusionAndFutureWork}
	\input{10-Acknowledgements}

	\bibliography{ncalbib}
	\bibliographystyle{unsrtnat}

	\input{A1-BigG}

\end{document}

%% file: 2-Introduction.tex
\section{Introduction}
\label{Introduction}

In recent years, gravitational wave physics has undergone a transformation from a search for a theoretical phenomenon to an observational science. The LIGO \citep{aLIGO} and Virgo \citep{virgo} gravitational wave observatories are now being used to study a growing number of astrophysical systems, including binary neutron stars \citep{GW170817, GW190425} and binary black holes \citep{GWTC, GW190412}.

In order to extract astrophysical source parameters from the measured signals, the observatory's readout systems must be precisely calibrated across the entire sensitive frequency band (20–2000 Hz) \citep{vitale2021physical, payne2020gravitational}. 
The accuracy of measured gravitational wave event parameters is crucial for astrophysical population studies \citep{abbott2020population}, cosmology \citep{ligo2017gravitational,abbott2021gravitational}, and probing for deviations from general relativity \citep{abbott2020tests}.
As such, substantial efforts have been made to yield calibration systematic errors which are less than $7\%$ in magnitude and $4$ degrees in phase for the LIGO and Virgo observatories \cite{O3Cal, VIR-0652B-19}.

A portion of reducing the systematic error relies on the accuracy and precision of the system's absolute reference, the Photon Calibrator (PCal) \citep{PCal,estevez2021advanced}.
These systems apply a direct, known force on the end test masses via radiation pressure.
Indirect options for absolute references have been used in the past, such as utilizing laser wavelength while in Michelson configurations \citep{FreeSwing, accadia2010virgo} and various laser frequency modulation techniques \citep{goetz2010calibration}.
However, the precision of these alternative methods has now been surpassed by the direct-force PCal systems \citep{goetz2010accurate, abbott2017calibration}.

Another promising direct-force absolute reference that has been previously discussed in the literature \citep{hirakawa1980dynamical, kuroda1985experimental, mio1987experimental, astone1991evaluation, astone1998experimental, Matone_2007} is applying a known gravitational force to the test masses with a system of rotating masses.
Such systems have recently been integrated at both the Virgo \citep{Estevez_2018, estevez2021newtonian} and KAGRA \citep{PhysRevD.98.022005} observatories. This paper describes the design, force modeling, first tests, and uncertainty budget of such a system for the LIGO observatories: the Newtonian Calibrator (NCal).

%% file: 3-Background.tex
\section{Background}
\label{Background}
A gravitational calibrator is an oscillating mass distribution that applies time-varying forces on a nearby test mass. 
This induces a known displacement of the test mass, and thus changes the differential arm length of the observatory.
These calibrators are usually designed as rotors with relatively simple geometries, which cause periodic changes in the local gravitational field as they rotate.

In the Newtonian limit, the gravitational force is the superposition of the forces caused by each of the rotor's elementary mass geometries - or equivalently, the mass multipole moments.
This allows one to design a rotor which produces forces at any multiple of the rotation frequency, $f$.
If a sub-geometry of the rotor has a mass distribution that is two-fold symmetric, it will produce a force at twice the rotation frequency, $2f$.
Likewise, a three-fold symmetric will produce a force at $3f$, etc.
Additionally, this allows one to disregard any sub-geometry whose mass distribution does not change with rotation.

Each term has a distinct dependence on the distance between the rotor and the test mass, $d$.
The $2f$ force caused by a two-fold symmetric quadrupole mass distribution will decrease proportional to $\sim1/d^4$.
While the force caused by a three-fold symmetric hexapole mass distribution ($3f$) will follow $\sim1/d^5$.
The force from higher multipole moments will fall off with higher orders of $d$.

For a fixed mass geometry, the force amplitude from each term is independent of rotation frequency.
Previously, rotors have been designed with geometries that produce signals at only $2f$ \citep{Matone_2007, Estevez_2018, estevez2021newtonian} and a mixture of both $2f$ and $3f$ \citep{PhysRevD.98.022005}.
The later systems have the advantage that they simultaneously inject forces at two frequencies.
Furthermore, due to the different distance dependencies, this design allows for $d$ to be measured from the observed ratio of the force amplitudes. \citep{PhysRevD.98.022005}.
Although the rotor described in this paper was designed with the goal of exploiting this fact, our surveying achieved a significantly higher precision measurement of $d$ than what could be reasonably achieved using the observed ratio \citep{T2000417,T2100066}.

%% file: 4-Hardware.tex
\section{Hardware}
\label{Hardware}

\begin{figure}[h!]
	\includegraphics[width=0.45\textwidth]{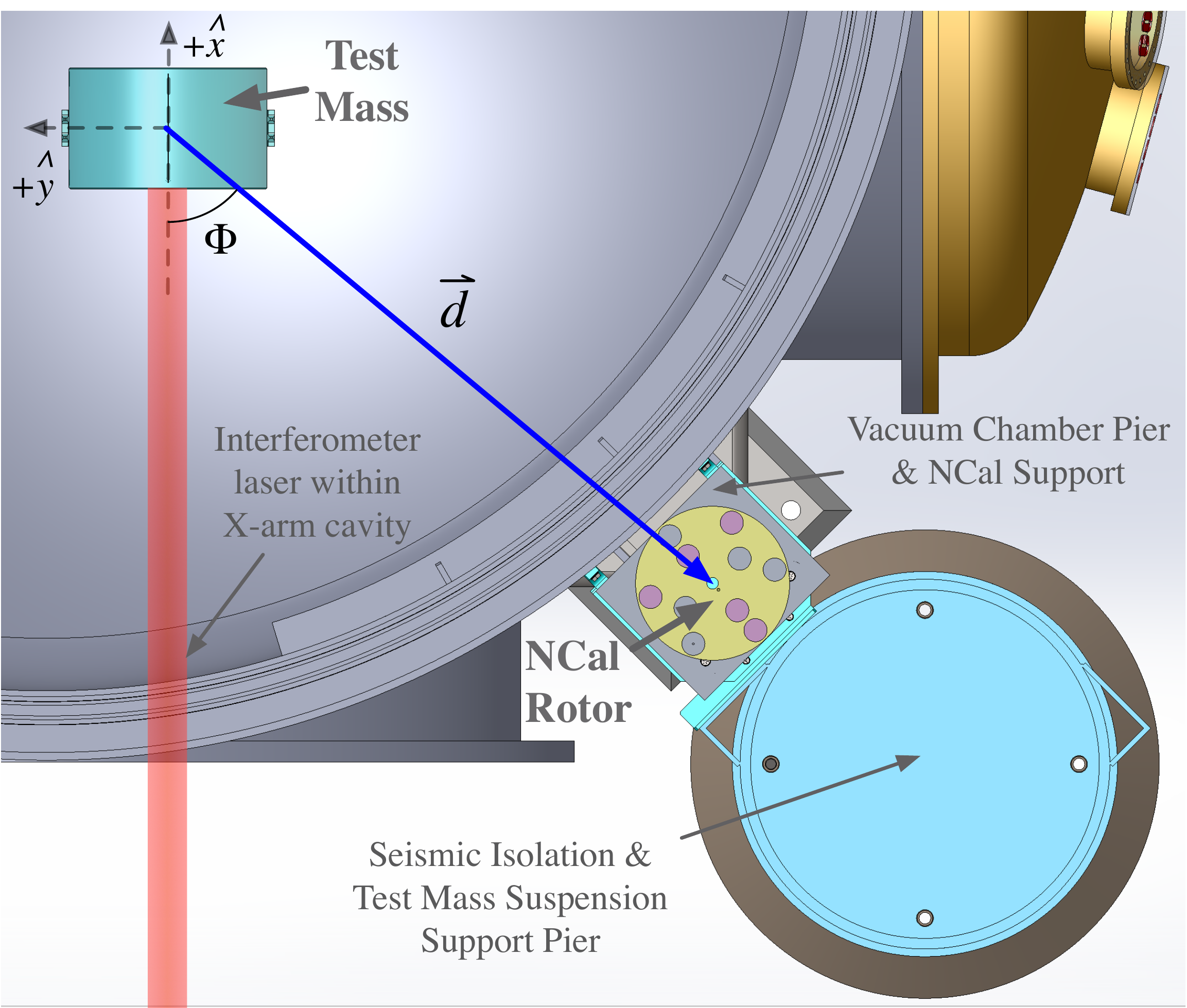}
	\caption{Top-down view of a CAD rendering of the NCal and test mass system.
		The NCal is shown without its enclosing frame, shell, or motor for clarity.
		The test mass is shown inside its large vacuum chamber with it's suspension and surrounding elements omitted. Also shown is the coordinate system used in this study along with the distance vector, $\vec{d}$, between the NCal and test mass. Note that the NCal rotor is slightly above of the $x$-$y$ plane.}\label{MountCAD}

\end{figure}

The NCal was installed at the X-end of the LIGO Hanford observatory during the commissioning break of LIGO's third observing run (O3).
The system was installed on a pier just outside the test mass vacuum chamber with a custom fabricated mounting structure, as shown in Figure \ref{MountCAD}.
This location was chosen as it was the closest in-air location to the test mass considering the pre-existing infrastructure.
The relevant dimensions of the system are detailed in the following sections and listed in Table~\ref{table:parameters}.

\subsection{Rotor}

\begin{figure}[h!]
	\centering
	\includegraphics[width=0.4\textwidth]{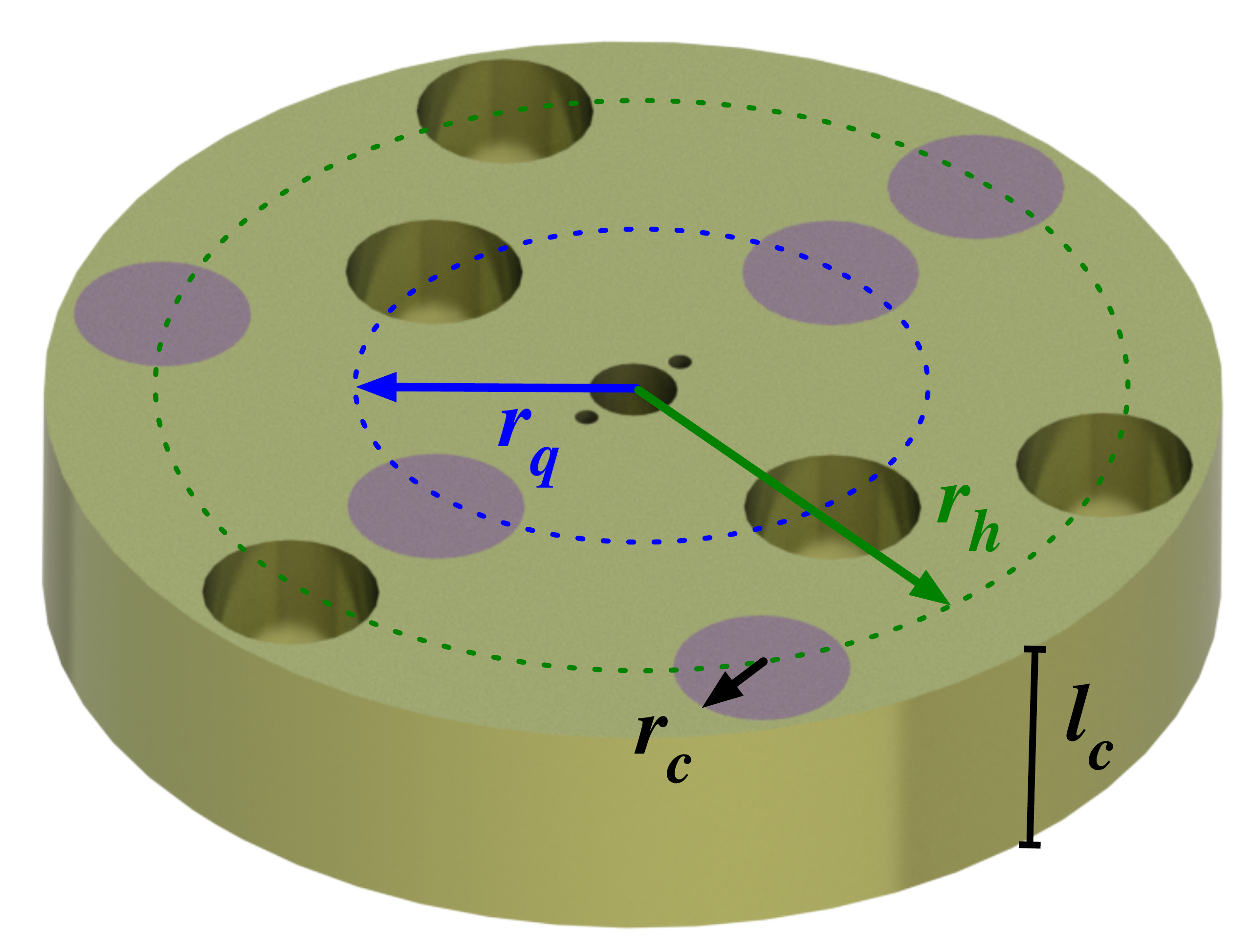}
	\caption{CAD rendering of the NCal rotor. Solid (purple) regions indicate the holes filled with tungsten cylinders and~gold regions indicating the aluminum disk and the unfilled cylindrically shaped holes. Geometrical parameters of the rotor are also highlighted for clarity.}\label{RotorCAD}
\end{figure}

The NCal rotor, shown in Figure \ref{RotorCAD}, consists of the two basic components.
The bulk of the rotor is an aluminum disk, 25.4~cm in diameter and 5.08~cm in height, with cylindrical cavities cut into it in four-fold and six-fold symmetric patterns.
A set of five tungsten cylinders, of radius $r_{c}$~=~1.968~cm and height of $l_{c}$~=~5.08~cm, form the primary~gravitating mass.
These were inserted into half of the disk's cavities to form two-fold and three-fold symmetric mass distributions at radii of $r_{q}~=~6.033$~cm  and $r_{h}~=~10.478$~cm, respectively.
A similar design was first proposed in \citet{Matone_2007} and developed for the KAGRA detector in \citet{PhysRevD.98.022005}.
The disk's cavities were milled to match the diameter of the cylinders in order to minimize differential motion between the cylinders and the disk.
The cylinders were cut from a single tungsten stock to minimize density differences.
The aspect ratio of the cylinders was chosen to decrease the harmonic content of the applied force and increase the simplicity of the force calculations.
Finally, the cylinders are restricted from moving out of the plane of the disk by an aluminum plate attached to the bottom of the disk.

\subsection{Test Mass}

The observatory's test masses are cylindrical-shaped fused-silica mirrors, suspended from a nearly identical penultimate mass via four 400 $\mu$m diameter, 0.57 m long fused-silica fibers \cite{cumming2012design}.
This monolithic system forms the last two stages of a quadruple pendulum system \cite{aston2012update}, which is suspended from a multi-stage seismic isolation system \cite{matichard2015seismic}.
The test mass is a cylinder of radius $r_{tm}$~=~170.11~mm with two vertical ``flats'' carved into opposite sides of the cylinder with a flat-to-flat width of $w_{tm}$~=~326.5~mm \cite{D080658}.
Fused-silica ``ears'' \cite{D080751} are bonded to these flats and welded to the fibers.
Further small, piezo-resistive tuned mass dampers are bonded to the test mass \cite{biscans2019suppressing}.
The bulk substrate of the Hanford End-X test mass was weighed in-air to have a mass of $39.611$~kg.
The two ears and four acoustic mode dampers add a total mass of $46.05$~g.
Additionally, the mass of the fiber below the effective bending point adds $0.7$~g.
The sum of this collection of massive elements, accounting for the $22$~g buoyancy correction, yields a mass of the test mass system of $M~=~39.680$~kg.

\subsection{Drive System}

\begin{figure}[h!]
	\includegraphics[width=0.4\textwidth]{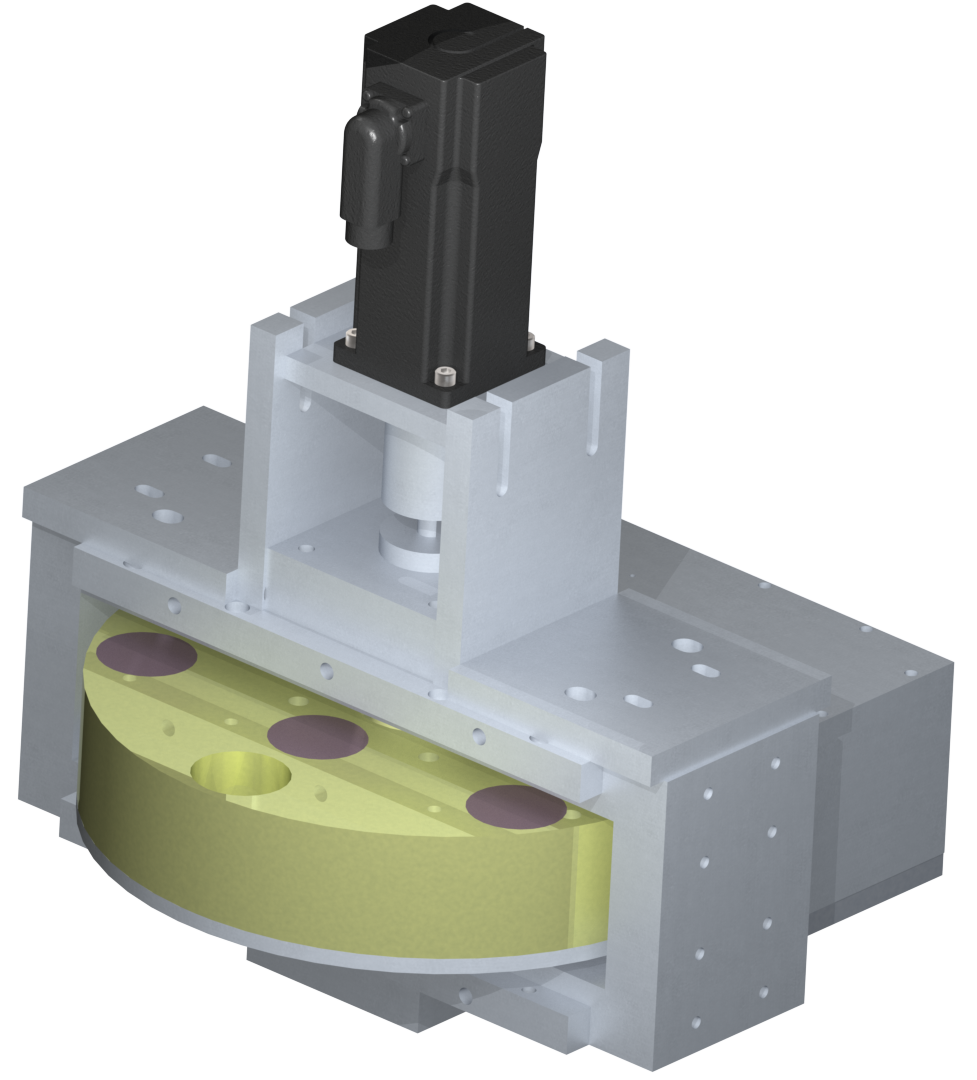}
	\caption{CAD rendering of the rotor, frame, and motor with the front cover omitted.}\label{HousingCAD}
\end{figure}

Figure \ref{HousingCAD} shows the rotor, frame, and motor of the NCal system.
The rotor is clamped to an aluminum shaft which runs through two sets of sealed ball bearings, one attached to the surrounding rigid frame above and below the rotor.
The bearings allow the rotor to spin freely while constraining the axis of rotation.

In order to decrease vibrations of the instrument, the aluminum plate was shifted to minimize the distance between the rotation axis and the rotor's center of mass. 
Additionally, the upper and lower bearings were tuned to align their rotation axes. 
These actions decreased the $1f$ vibration to an acceptable amplitude of $<0.5\ \text{m/s}^2$ at a rotation rate of 27.7 Hz.

A three-phase, synchronous motor is coupled to the top of the shaft and was driven by a commercially-available Beckhoff motor controller.
The rotation rate was locked in feedback with the motor's internal rotary encoder.
This system kept the rotation frequency constant to within $<0.1\%$.
A secondary rotary encoder was connected to the bottom of the shaft which independently monitored the rotation frequency. 
Unfortunately, the absolute phase between the encoders and the rotor masses was not aligned during installation which restricted our ability to predict the phase of the NCal forces.

%% file: 6-Modelling.tex
\section{Models}\label{modelling}

As previously outlined in Section \ref{Background}, the net force acting on the test mass can be written as the sum of the forces caused by the sub-geometries which make up the rotor.
The force from our rotor can be broken into five terms:
\begin{eqnarray}
    \vec{F} & = & \vec{F}_{\text{Al Disk}}^{(0f)}+\vec{F}_{\text{Quad. Cyl.}}^{(2f)}+\vec{F}_{\text{Hex. Cyl.}}^{(3f)} \nonumber\\
    & &\hspace{2.0cm}~+~  \vec{F}_{\text{Oct. Holes}}^{(4f)}+\vec{F}_{\text{Dodec. Holes}}^{(6f)} \label{eq:fullforcevector}
\end{eqnarray}
where the subscripts denote the sub-geometry, and the superscripts indicate the lowest frequency at which each term contributes.
The first term is the time-independent force due to the aluminum disk, the second and third terms are due to the tungsten cylinder masses that are, respectively, in quadrupole and hexapole configurations, and the last two terms are due to the holes in the aluminum disk which form both an octupole and a dodecapole.
The contributions due to the holes can be treated as negative mass when calculating the force.

Since the observatory is insensitive to time independent forces, we can disregard the aluminum disk term.
Due to the large density difference between tungsten (19.3 g/cm$^3$) and aluminum (2.7 g/cm$^3$), the NCal primarily produces forces at $2f$ and $3f$.
The missing mass of the holes also produces forces at $4f$ and $6f$.
However, these are subdominant due to both the extra distance attenuation of these moments and the decreased effective mass of the holes.
Additionally, the geometry inherently has higher moment contributions which are even weaker than the $4f$ and $6f$ forces.
While our modeling methods are capable of accounting for these higher moments, no frequency content higher than $3f$ was observed in the measurements described in this paper.
Thus only models (and subsequent comparison with measurement) for the $2f$ and $3f$ forces are discussed here.

We used three independent models to calculate the expected gravitational force between the NCal and the test mass.
First, we used a point mass approximation to derive equations that can be understood analytically and quickly evaluated.
To account for the full geometry of the system, we developed two other models: a finite-element simulation, and a multipole method analysis.
We employed Monte Carlo methods to estimate the uncertainty on the predicted force amplitudes for each model.
This begins by computing the force at a fixed orientation of the rotor with a collection of input parameters randomly sampled from their respective probability distributions.
The force is then recomputed at a collection of rotor rotation angles to represent the time evolution of the force.
The $2f$ and $3f$ amplitudes are then extracted by fitting to the corresponding sinusoidal functions.
This is then repeated with a new randomly sampled set of parameters to yield the distributions shown in Figure \ref{ForceDist}.
This method inherently accounts for all degeneracies and non-linearities in the force calculations.

\begin{widetext}
\begingroup
\setlength{\tabcolsep}{10pt} 
\renewcommand{\arraystretch}{1.5} 

\begin{table}[h!]
\begin{center}
\begin{tabular}{ |l|c|c|c| }
\hline
 Parameter & Mean & Uncertainty & Distribution\\
 \hline
Tungsten Cylinder Mass $m$ & 1.0558~kg & 0.3~g &~Gaussian \\
Tungsten Cylinder Radius $r_c$& 1.968~cm & 2.5 $\mu$m &  Uniform\\
Tungsten Cylinder Length $l_c$& 5.08~cm & 5 $\mu$m & Uniform \\
Quadrupole Radius $r_{q}$ & 6.033~cm & 5 $\mu$m & Uniform\\
Hexapole Radius $r_{h}$ & 10.478~cm & 5 $\mu$m & Uniform \\
Test Mass $M$ & 39.680~kg & 10~g & Uniform\\
Test Mass Length $l_{tm}$& 199.8~mm & 0.1~mm & Uniform\\
Test Mass Radius $r_{tm}$& 170.11~mm & 0.05~mm& Uniform \\
Test Mass Flat Width $w_{tm}$& 326.5~mm & 0.05~mm& Uniform \\
Rotor Radial Position $\rho$ & 1180.2~mm & 2.4~mm & Numerical\\
Rotor Angle $\Phi$ & 52.24 deg & 0.08 deg& Numerical\\
Rotor Vertical Position $z$ & 10.0~mm & 3.0~mm & Numerical\\
 \hline

 \end{tabular}
 \caption{Parameters of the NCal rotor and LIGO test mass, their uncertainties, and their corresponding distribution types.
 	For parameter uncertainties described by a~gaussian distribution the stated uncertainty indicates the $\sigma$-value; for uniform it indicates the half-width; and for uncertainties estimated numerically, the 68\% credible interval -- see further discussion of uncertainty in Section \ref{subsec:uncert}.}\label{table:parameters}
 \end{center}

\end{table}
\endgroup
\end{widetext}

With the NCal installed near the End-X test mass, only the $x$-component of the force is observed by the observatory.
The point mass approximation yields the $x$-component time series for the $2f$ and $3f$ forces directly while the full geometry models provide time series containing the full harmonic content of the force in all three directions.
For these methods, the $2f$ and $3f$ force amplitudes are extracted by projecting the force onto the $x$-axis.

The two full geometry methods treat the tungsten cylinder and the holes in the aluminum disk as perfect cylinders and the test mass as a cylinder with flat sides.
The aluminum disk is omitted as it does not produce a time-varying force. Additionally, the ears and suspension fibers are omitted as their geometric contributions are negligible at the current precision.

 \subsection{Point Mass Approximation}

To find the functional form of the force, we derived analytical expressions of the $2f$ and $3f$ components by approximating each tungsten mass and the test mass as point masses at their respective centers of mass.
The analytical methods used in this approximation were originally developed in \citet{Matone_2007} and furthered in \citet{Estevez_2018} for a quadrupole mass arrangement.
Here, we have extended the method to include the hexapole mass arrangement.
The quadrupole representation shown here is numerically identical to the result from \citet{Estevez_2018}. The derivation of the following equations are provided in \citet{T2000238}.

With the assumption that the rotor is nearly in-plane with the test mass ($z/d\ll 1$, $\rho \approx d$), the $x$-component of the quadrupole force is:

\begin{eqnarray}
F_{x}^{(2f)}& = & \frac{9}{2}~\frac{G M m r_{q}^2}{d^{4}} \nonumber\\
& & \hspace{0.5cm} ~\times~ \zeta^{-5/2} \bigg[\left(\frac{5}{6}\frac{1}{\zeta} - \frac{2}{3} \right) \cos(2\theta - \Phi) \nonumber\\
& & \hspace{2.2cm} ~+~ \frac{5}{6}\frac{1}{\zeta} \cos(2\theta - 3\Phi) \bigg],
	\end{eqnarray}
where $G$ is the gravitational constant, $M$ is the mass of the test mass, $m$ is the mass of a tungsten cylinder, $d$ is the distance between the rotor and test mass, $\Phi$ and $z$ are the azimuthal and vertical position of the rotor (as in Figure \ref{MountCAD}), $r_q$ is the quadrupole radius, and $ \zeta = 1 + (z/d)^{2}$. The angle of the rotor, $\theta$, cycles once per rotation (i.e. $\theta=2 \pi f t$).

Under the same assumptions, the $x$-component of the hexapole force is:

\begin{eqnarray}
F_{x}^{(3f)}& = & \frac{15}{2}\frac{G M m r_h^3}{d^{5}}\nonumber\\
& & \hspace{0.5cm} ~\times~ \zeta^{-7/2} \Bigg[\left(\frac{7}{8}\frac{1}{\zeta} - \frac{3}{4}\right) \cos(3\theta - 2\Phi) \nonumber\\
& & \hspace{2.2cm} ~+~ \frac{7}{8}\frac{1}{\zeta} \cos(3\theta - 4\Phi) \Bigg]
\end{eqnarray}
where $r_h$ is the radius of the hexapole mass arrangement. These equations are the results of expansions to second and third order in $r_{q}/d$ and $r_{h}/d$, respectively.

While these analytical expressions are approximate, they provide an consistency check of the full-geometry models and can be rapidly computed.

\subsection{Finite-Element Simulation}\label{sim}

The first full-geometry method we used is to numerically compute the force by approximating the geometry as two clouds of finite-element point masses.
The force at a given $\theta$ is then simply the sum of all forces between pairs of particles, one in the ``source'' cloud and the other in the ``test'' cloud:

\begin{equation}\label{equation_pointmass}
    \vec{F}=\sum_i \sum_j \frac{G m_i M_j}{d_{ij}^2}\ \hat{d}_{ij}
\end{equation}
where $m_i$ is a point mass within the source cloud (the NCal), $M_j$ is a point mass within the test cloud (the test mass), and $d_{ij}$ is the magnitude of the position vector from the $i$-th point to the $j$-th point.

A code utilizing the \textit{PointGravity} algorithms of the \texttt{newt} libraries \cite{Hagedorn, pgURL} was developed which provides the force in all three directions, as well as calculating the torques acting on the test mass about all three axes.

For each object that makes up the point clouds (tungsten cylinder or test mass), the \textit{PointGravity} algorithm creates a Cartesian lattice of point masses with the minimum dimensions needed to contain the object.
The points within the cylindrical volume of the given object are then selected out of this lattice and renormalized to account for the mass of the lost points.
We chose to have equal number of lattice points in each dimension and, through a convergence test, found that a lattice point number of 10 resulted in an acceptable inaccuracy of $<0.01 \%$.

\subsection{Multipole Analysis}

In addition to the finite-element simulation, we developed a multipole-based analysis which calculates the force by breaking each object into its gravitational multipole moments.
With this method, the force \cite{Force} can be found with:
\begin{equation}
    \vec{F}=4\pi G \sum^{\infty}_{l=0} \sum^{l}_{m=-l} \frac{1}{2l+1} Q_{lm} \nabla q_{lm} \label{Fmulti}
\end{equation}
where $Q_{lm}$ is the ``outer'' multipole and $q_{lm}$ is the ``inner'' multipole. For our geometry the inner multipole corresponds to the NCal while the outer corresponds to the test mass.

The forces along all three axes were calculated using the \textit{Multipole} algorithms of the {\tt newt} libraries \cite{pgURL}. These libraries construct the system's geometry from a collection of elementary solids whose multipole moments have been analytically determined \cite{Hoyle_2006, Stirling}. The forces between the rotor and the test mass are then calculated using methods based on a wide collection of literature sources \cite{Force, Trenkel, DUrso}.

To numerically evaluate the force using Equation \ref{Fmulti}, an upper cutoff for the $l$-sum must be imposed. For our analysis, this was chosen to be $l=11$. Moments higher than $l=11$ only contribute significantly to frequencies greater than $11f$.

\subsection{Parameter Uncertainty}\label{subsec:uncert}

The estimated force and uncertainty for each model is determined numerically using Monte Carlo methods with parameters defined by a selection of probability distributions.
The distribution type and defining characteristics are listed in Table \ref{table:parameters} and motivated below.

Geometric parameters of the cylinders, $r_{c}$, $l_{c}$, $r_{q}$, and $r_{h}$, were determined with a single micrometer measurement with a known precision.
For these parameters, we assigned a uniform distribution centered on the measured value and bounded by the precision.
Geometric parameters of the test mass, $l_{tm}$, $r_{tm}$, and $w_{tm}$ are determined from tightly controlled  vendor specifications and are also assigned uniform probability distributions bounded by their respective tolerances.

The force models assume an equal mass, $m$, for all five tungsten cylinders.
We take $m$ to be a Gaussian distribution with a central value equal to the mean of the measured masses, and a $\sigma$-value equal to their standard deviation.
In contrast, since there is only one test mass, we set the distribution for $M$ in a similar way as geometric parameters of the cylinders: a uniform probability distribution bounded by the estimated precision.

Surveying equipment was used to determine the position of the NCal frame and the front surface of test mass with respect to monuments located on the floor and walls of the observatory.
For the NCal, the frame location was combined with a CAD model confirmed by bench-top measurements to give the position of the rotor's center of mass.
Similarly, the position of the test mass center of mass was found by shifting the $x$-position of the front surface by half it's thickness and the $z$-position was altered to account for lack of buoyancy in vacuum.
Finally, these two positions were subtracted to yield the position, $\rho$, $\Phi$, and $z$, of the NCal in coordinates centered on the test mass center of mass.
As the process involved several non-linear transformations, the probability distribution of the relative position was calculated numerically through Monte Carlo methods \citep{T2000417}.

\begin{widetext}

	\begin{center}

		\begin{figure}[!h]

			\includegraphics[width=0.85\textwidth]{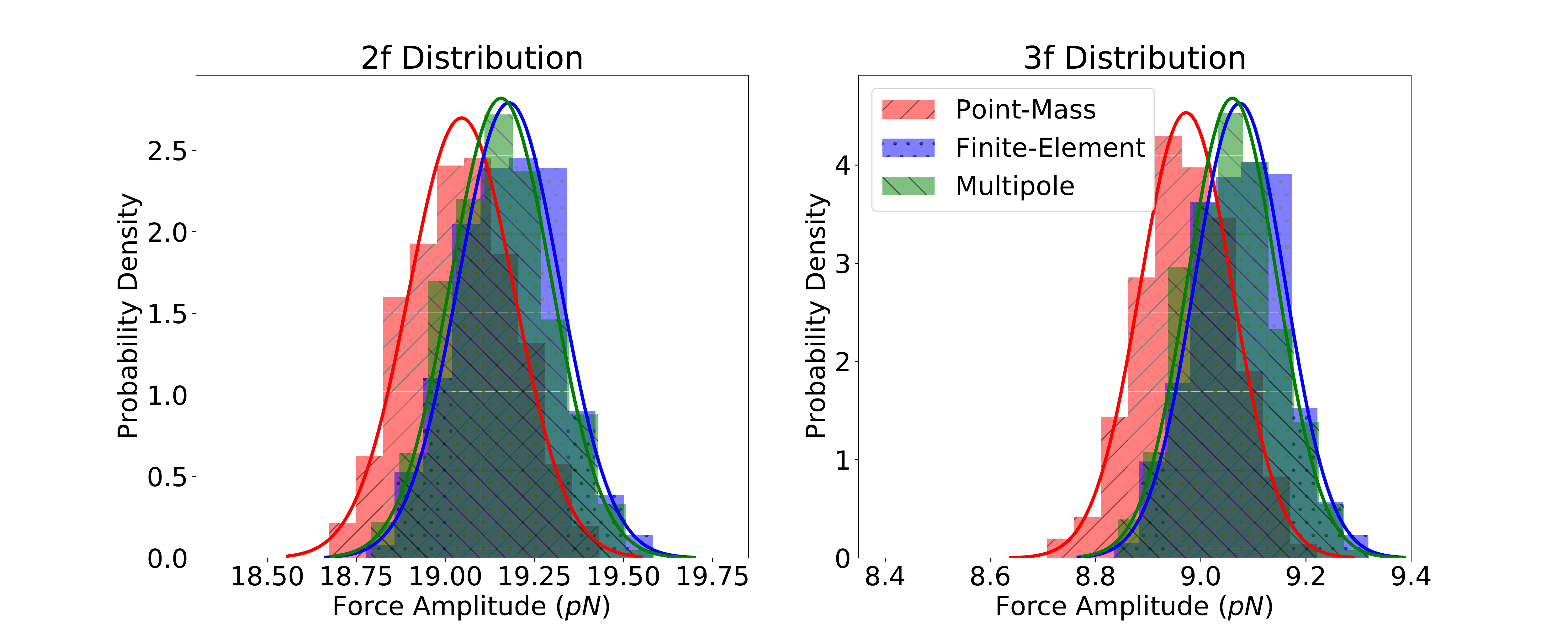}

			\caption{Distributions of the predicted force amplitudes, $F_{x}^{(2f)}$ and $F_{x}^{(3f)}$, calculated by the point-mass, finite-element, and multipole models as well as Gaussian fits to each distribution. The small discrepancy between the point-mass model and the other two models is due to it lacking the full geometry of the system.}\label{ForceDist}
		\end{figure}

	\end{center}

\end{widetext}

\subsection{Model Comparison}
Histograms of the distribution of force amplitudes from each model are shown in Figure \ref{ForceDist}. The distributions are computed from a set of 800 samples for each model. The number of samples was chosen to maximize the accuracy of the distribution while minimizing the computational cost for the finite-element analysis.
Each model's distribution is fit to a Gaussian whose mean and $\sigma$-value is taken to be the estimated value and uncertainty, respectively, and are reported in Table~\ref{PredictedRotationTable}.

\begingroup
\setlength{\tabcolsep}{12pt} 
\renewcommand{\arraystretch}{1.5} 
\begin{table}[!ht]
\begin{center}
\begin{tabular}{ |c|c|c|}
 \hline
 Model & $F_{x}^{(2f)}$ (pN)& $F_{x}^{(3f)}$ (pN) \\
 \hline
Finite-element & $19.18^{\pm 0.14}_{(\pm 0.75\%)}$& $9.07^{\pm 0.09}_{(\pm0.95\%)}$ \\
Multipole &$ 19.16^{\pm 0.14}_{(\pm 0.74\%)}$& $9.06^{\pm 0.09}_{(\pm 0.94\%)}$  \\
Point-mass & $19.04^{\pm 0.15}_{(\pm 0.76\%)}$ & $8.97^{\pm 0.09}_{(\pm0.95\%)}$  \\
\hline
\end{tabular}
\caption{Predicted force amplitudes and uncertainties extracted from the distributions shown in Figure \ref{ForceDist} for the finite-element, multipole, and point-mass models}\label{PredictedRotationTable}
\end{center}
\end{table}
\endgroup

All three models provide almost identical levels of uncertainty at $\sim0.75\%$ and $\sim0.95\%$ for the $2f$ and $3f$ force amplitudes, respectively.
However, the point mass approximation predicts a lower mean value than the other two models.
This discrepancy is due to the point mass approximation's improper treatment of the geometry of the system.
While the finite-element model accurately captures the full system, we take the multipole model to be the most robust model as it captures the entire geometry of the system and utilizes exact solutions for the multipole moments.
As such, we use the results of the multipole analysis to compare with measurement while the other two models are only used to verify the accuracy of the multipole force predictions.

%% file: 7-Measurements.tex
\section{Measurements}
\label{Meas}

On December 4, 2019, gravitational forces were injected with the NCal while the observatory was in its observation-ready noise state. 
The rotor rotation frequency, $f$, was stepped through five values ranging from 4 to 10 Hz which injected forces between 8 to 30 Hz. 
Explicit values for $f$, $2f$, and $3f$ are listed in Table \ref{RotationTable}.
Investigations into the impact of the NCal electronics in addition to the vibrational noise contributions found no significant increase in the observatory's noise performance during the operation of the NCal.	
For further discussion of extraneous couplings, see Section \ref{subsec:extracouplings}.

The observatory's differential strain readout, $h$, was recorded continuously during the NCal injections.
Any observed strain at the expected $2f$ and $3f$ frequencies was attributed to the force by the NCal on the test mass.
As such, we convert the observed strain into force in the $x$-direction with:
\begin{eqnarray}
	F_{x} & = & L\ h~/~S \label{eq:dF_ASD}
\end{eqnarray} 
where $h$ is the strain output, $L$ is the arm length, and $S$ is the magnitude of the force-to-displacement response of the test mass suspension system.

\begin{widetext}

\begin{center}

\begin{figure}[!h]
	\includegraphics[width=0.90\textwidth]{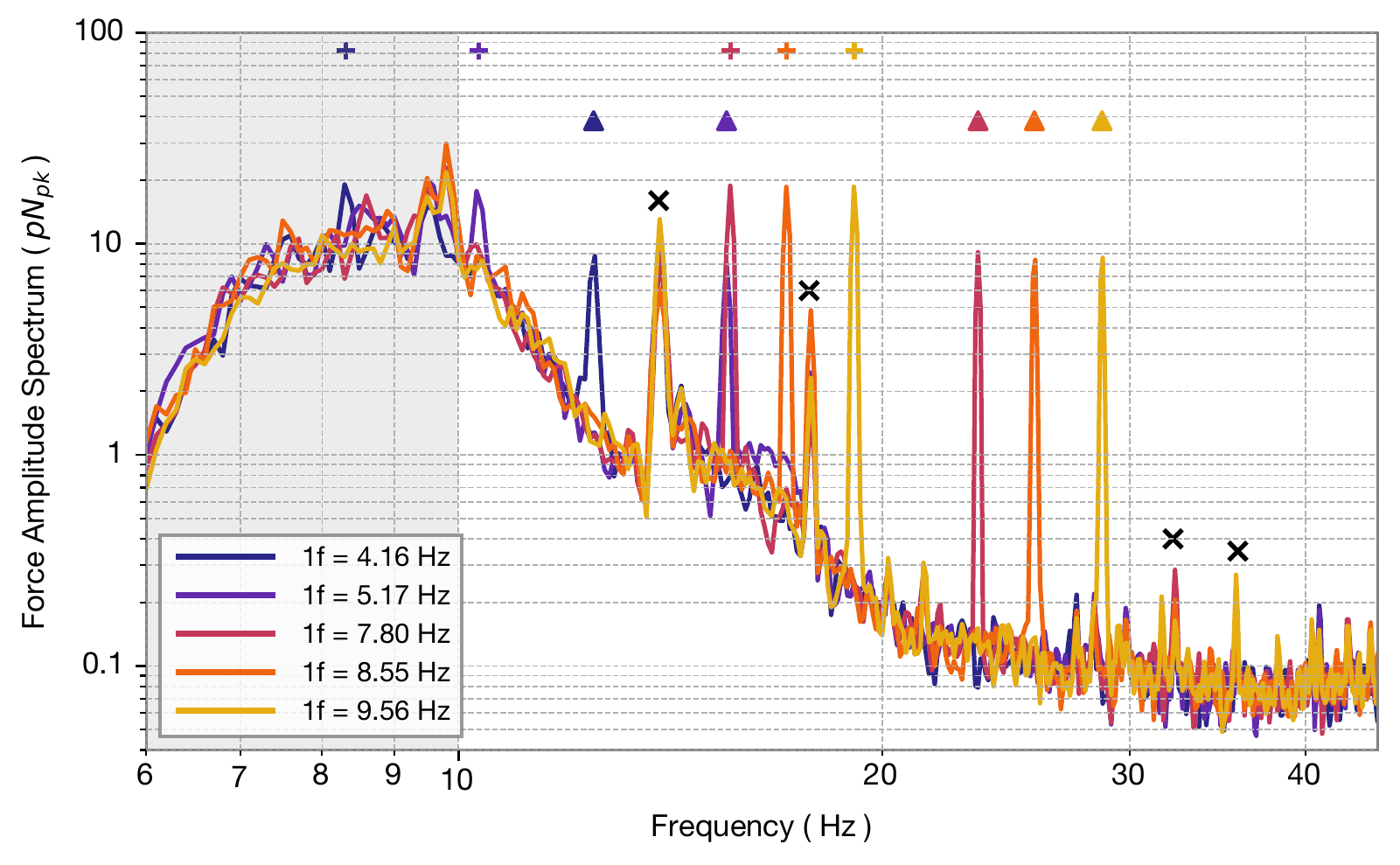}
	\caption{The collection of amplitude spectra of the observed force during successive NCal injections. Each colored trace represents a different injection when the NCal was rotating at the corresponding $1f$ rotation frequency. As visual aides, similarly colored plus- and triangle-shaped markers indicate the location of the $2f$ and $3f$ frequencies at arbitrary amplitude. Black cross-shaped markers highlight ever-present spectral features that are unrelated to the NCal injection. The gray shaded region indicates the frequency band where the strain readout is aggressively high-passed and thus should be ignored.}\label{ForceMeas}
\end{figure}

\end{center}

\end{widetext}

Figure \ref{ForceMeas} shows the collection of amplitude spectra of the force on the test mass during the injections.
Each overlaid color amplitude spectra corresponds to a different rotor rotation frequency.
Although these spectra are not used in our results, they convey the sensitivity of the observatory during the NCal injections.
The $2f$ and $3f$ forces are clearly resolved for most rotation frequencies and the signal-to-noise improves with frequency. 
The $3f$ signal from $f = 5.17$ Hz and $2f$ signal from $f = 7.80$ Hz intentionally overlap.
No additional features or broad-band noise were observed during these injections.
The $4f$ and $6f$ forces caused by the holes in the rotor may be resolved with future, long duration injections.
Artifacts, unrelated to the NCal injections, can also be seen in Figure \ref{ForceMeas}. 
First, the strain below 10 Hz is to be ignored as it is aggressively high-passed to aide astrophysical search algorithms. 
Additionally, there are ever-present spectral features that must be avoided when choosing injection frequencies.

The force amplitudes are extracted using least squares spectral analysis \citep{T2100103}. 
This consists of fitting the strain readout in the time-domain using linear least squares to the following sinusoidal function:
\begin{align}
h(t)=&a_2 \cos(2\pi\ 2 f\ t)+b_2 \sin(2\pi\ 2 f\ t)\nonumber \\
&+a_3 \cos(2\pi\ 3 f\ t)+b_3 \sin(2\pi\ 3 f\ t)
\end{align}
where $a_i$ and $b_i$ are the cosine and sine amplitudes at the respective frequencies, $f$ is the rotor rotation frequency, and $t$ is time.
The data was cut into sections with lengths equal to 20 $1f$-periods. 
Each cut was then fit independently and the extracted sine and cosine amplitudes were averaged before being combined in quadrature to yield a strain amplitude:
\begin{align}
h(f_i)=\sqrt{a_i^2+b_i^2}
\end{align}
Finally, the extracted strain amplitudes were then converted to force using Eq. \ref{eq:dF_ASD}. 
The uncertainty on the observed force amplitude was calculated by propagating the standard deviation of the sine and cosine amplitudes through Eq. \ref{eq:dF_ASD}. 

The least squares spectral analysis method has multiple benefits over a traditional Fourier-based spectral analysis.
First, it guarantees that the amplitude is extracted for the frequency of interest (i.e. frequency bins are centered on the frequencies of interest). 
Additionally, the extracted sine and cosine amplitudes are guaranteed to be Gaussian distributed as long as the measured strain is also Gaussian. 
The statistical uncertainty can also be extracted directly from the observed distribution of amplitudes. 
Lastly, there are no correction factors that need be included which decreases the nuance of the analysis and removes possible sources of error.

%% file: 8-Results.tex
\section{Systematics and Total Uncertainty}
\label{subsec:extracouplings}

In addition to the gravitational force from the NCal, the observed force is influenced by a variety of effects that must be accounted for to accurately extract the measured force.
These effects include indirect gravitational coupling, non-gravitational coupling, and distortion due to the observatory's strain readout.
These influences can be taken into account in a few ways: by adding a correction to either the measurement or the model, or by expanding the uncertainty in either to account for these contributions.
Due to the nature of these systematics and extraneous couplings, we chose to correct and expand the uncertainty of the measurement.
The corrections and uncertainties for each systematic are shown in Table~\ref{uncert} along with the measurement and model uncertainties.

\subsection{Indirect gravitational coupling}

The NCal rotor not only gravitationally attracts the test mass but also pulls on the rest of the suspension system.
The penultimate mass is the next closest component of the suspension, 602~mm above the test mass center of mass \cite{D0901346}, and transmits motion to the test mass through the suspension fibers.
This coupling path was studied using the finite-element simulation described in Section~\ref{sim}.
This simulation found that -- aside from the quadrupole influence at 10.34~Hz -- the penultimate contributions are less than $0.1\%$ of the expected forces \citep{T2100095}. 
Even at 10.34~Hz, the coupling increases to only $\sim0.17\%$.
Since this is much less than the measurement statistical uncertainty, we expanded each measurement uncertainty by the corresponding penultimate mass contribution. 
This contribution is labeled in Table \ref{uncert} as Sys: Pen.
The forces on the rest of the suspension are expected to be negligible due to the their increased distance from the NCal rotor.

\begingroup
\setlength{\tabcolsep}{10pt} 
\renewcommand{\arraystretch}{1.25} 
\begin{center}	
\begin{table}[!h]
	\begin{center}
		\begin{tabular}{ |l|c|c|}
			\hline
			Source & \specialcell{Correction (pN), \\ $[$Ratio Relative\\ to Model, (\%)$]$} & \specialcell{Uncertainty (pN), \\ $[$Ratio Relative\\ to Model, (\%)$]$} \\
			\hline
			\multicolumn{3}{|c|}{} \\ [-12pt] 
			\hline
			\multicolumn{3}{|c|}{$2f$}\\
			\hline  
			\multicolumn{3}{|c|}{} \\ [-12pt]
			\hline
			\specialcell{Model: $F_{x}^{(2f)}$}  & -- & \specialcell{0.14,\\$[0.74]$}\\
			\hline
			\specialcell{Meas: Stat.} & -- & \specialcell{0.21-2.33\\$[1.09-12.16]$}\\ \hline
			\specialcell{Sys: Pen.} & -- & \specialcell{0.007-0.033\\$[0.04-0.17]$}\\  \hline
			\specialcell{Sys: Mag.} & -- & \specialcell{0.04\\$[0.21]$} \\  \hline
			\specialcell{Sys: $S$} & -- & \specialcell{0.022-0.064\\$[0.11-0.34]$} \\ \hline
			\specialcell{Sys: Torque} & \specialcell{0.339\\$[1.77]$} &\specialcell{0.026\\$[0.14]$}\\  \hline
			\multicolumn{3}{|c|}{} \\ [-11.5pt] \hline
			Meas \& Sys & -- &\specialcell{0.22-2.33\\$[1.12-12.17$]} \\ \hline
			\multicolumn{3}{|c|}{} \\ [-12pt] \hline
			\multicolumn{3}{|c|}{$3f$}\\ \hline
			\multicolumn{3}{|c|}{} \\ [-12pt] \hline
			\specialcell{Model: $F_{x}^{(3f)}$} & -- & \specialcell{0.09\\$[0.94]$}\\ \hline
			\specialcell{Meas: Stat.} & -- & \specialcell{0.24-0.57\\$[2.64-6.31]$}\\ \hline
			\specialcell{Sys: Pen.} & -- & \specialcell{0.0002-0.0074\\$[0.002-0.081]$} \\ \hline
			\specialcell{Sys: Mag.} & -- & \specialcell{0.04\\$[0.44]$} \\ \hline
			\specialcell{Sys: $S$} & -- & \specialcell{0.005-0.024\\$[0.05-0.26]$}\\ \hline
			\specialcell{Sys: Torque} & \specialcell{0.199\\$[2.19]$}& \specialcell{0.015\\$[0.17]$}\\ \hline
			\multicolumn{3}{|c|}{} \\ [-12pt]\hline
			Meas \& Sys & -- &\specialcell{0.24-0.57\\$[2.69-6.34]$}\\ \hline
		\end{tabular}
		\caption{Comparison between the amplitude uncertainty in the model (Model: $F_{x}^{(2f)}$ and $F_{x}^{(3f)}$), the measured force (Meas: Stat.), the various systematics (Sys), and the total measured and systematic uncertainty (Meas \& Sys). The ratio between the contributions and the multipole model are shown in square braces in percent. Contributions which are frequency-dependent are shown as a range between their minimun-maximum values.}\label{uncert}
	\end{center}
\end{table}
\end{center}
\endgroup

\subsection{Non-gravitational coupling}
One concern that may skew the measured force is that the NCal rotor and test mass could be coupled through magnetic effects.
Using existing magnetic coupling measurements for the test mass \citep{nguyen2021environmental} and a collocated magnetometer, we constrained the force due to magnetic coupling to be $<0.04$ pN across the frequency band of interest. \citep{T2100221}
To account for this, we add a conservative uncertainty of 0.04 pN to each measured force.
This contribution is labeled as Sys: Mag in Table \ref{uncert}.
With dedicated magnetic coupling measurements and mitigation, we expect this contribution to decrease for future studies.

Another non-gravitational coupling that may influence the measurements is the transmission of vibrations caused by the NCal.
Vibrations caused by the NCal rotor can couple into the strain measurements through the seismic isolation system as well as through scattered light.
The frequency region of existing vibrational coupling measurements \citep{nguyen2021environmental} do not extend to our band of interest.
As such, we were unable to constrain the vibrational contribution to the observed force. 
However, accelerometers located near the rotor did not observe any change in the ambient vibrations while the NCal was rotating \citep{T2100221}.
Since ambient vibrations are known to not contribute strongly to the observatory's noise level \citep{nguyen2021environmental}, we expect that vibrations are not a significant source of the observe force.
This will be studied in detail with future dedicated vibration coupling injections and measurements.

\subsection{Detector readout systematics}
In addition to forces on the suspension system, the NCal rotor applies torques to the masses.
Again due to the increased distance dependence of the gravitational interactions, the torques on the test mass are expected to be significantly larger than those on the rest of the suspension system.
These torques couple into the strain measurements due to the interferometric beam being offset from the center of rotation of the test mass. 
During O3, the beam was offset by $13.2\pm 1$~mm in the $y$-direction and $-15.7\pm 1$~mm in the $z$-direction.
\begin{widetext}

\begin{center}
	\begin{figure}[!h]
		
		\begin{subfigure}[t]{0.85\textwidth}
			\includegraphics[width=\textwidth]{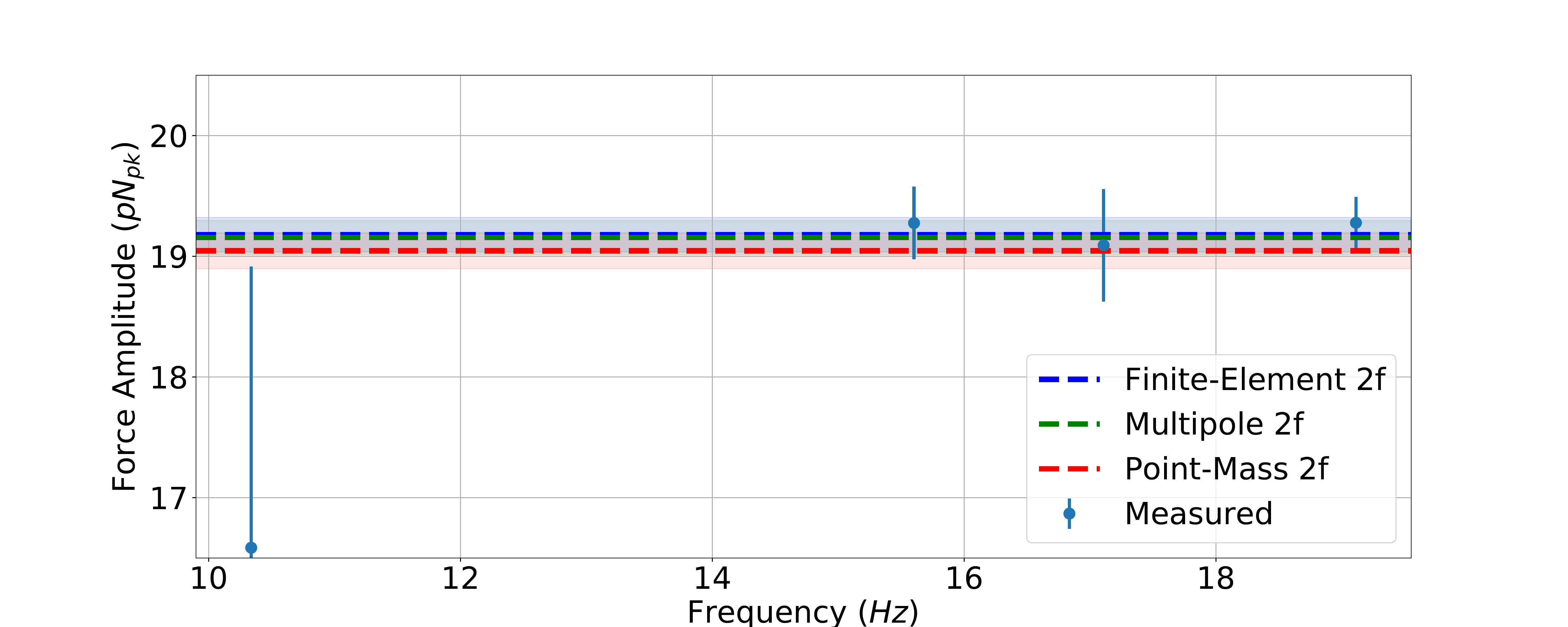}
			\caption{Force exerted by the quadrupole mass arrangement, $F_{x}^{(2f)}$, as measured by the detector compared with the frequency-independent values predicted by the models from Section \ref{modelling}.}\label{2Fcomp}
		\end{subfigure}
		
		\begin{subfigure}[t]{0.85\textwidth}
			\includegraphics[width=\textwidth]{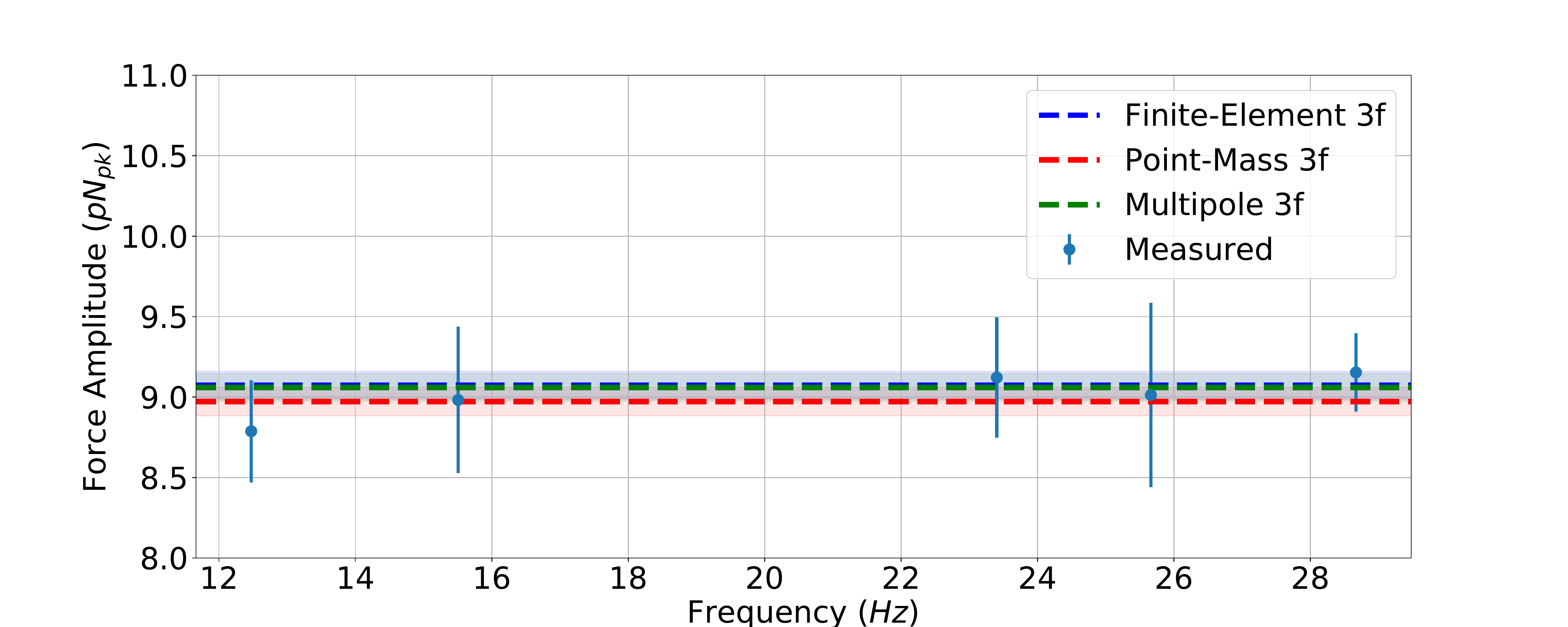}
			\caption{Force exerted by the hexapole mass arrangement, $F_{x}^{(3f)}$, as measured by the detector compared with the frequency-independent values predicted by the models from Section \ref{modelling}.}\label{3Fcomp}
		\end{subfigure}
		\caption{}\label{comp}
	\end{figure}
\end{center}

\end{widetext}	
This offset causes the angle change due to torques about the $y$ and $z$ axes (pitch and yaw) to cause arm length changes.
This arm length change is then interpreted as force in our measurements.
The torque to force coupling was also studied using the finite-element simulation described in Section \ref{sim} \citep{T2100088}. The apparent force due to both torques summed together is $0.339\pm0.026$ pN for the $2f$ and $0.199\pm0.015$ pN for the $3f$.
We shift each measured force by the correspond amount, accounting for relative phase, to remove this contribution and add the uncertainty to each measurement point.
This shift is the only correction we apply as its phase is well-known from the model.
The corrections and corresponding uncertainties are listed in Table \ref{uncert} as Sys: Torque.
The torques on the rest of the suspension chain are expected to be negligible as they will be attenuated by the extra distance.

As discussed in Eq. \ref{eq:dF_ASD}, the strain readout is converted to force via the test mass force-to-displacement transfer function $S$.
Above 10 Hz, we approximate the test mass as a free mass, $S~=~-1/(M (2\pi f)^2)$.
Comparing this to a full model of the complete dynamics of the suspension shows this adds a frequency-dependent error of no more than $0.4\%$ above 10 Hz \cite{Bhattacharjee_2020}.
The uncertainty of the measured force has been expanded to encompass this expected error and is listed in Table \ref{uncert} as Sys:~$S$.

\subsection{Results \& Total Uncertainty}

The measured force amplitudes are listed in Table \ref{RotationTable} and shown in Figure \ref{comp} alongside the predictions from each model described in Section \ref{modelling}.
The measured force is consistent with the frequency-independent predictions at all frequencies with the only excursion being at 10.34 Hz, where the detector noise is worse and many of the systematics have their largest effect.
Due to the limited time spans of the injections, the statistical uncertainty for every measurement is larger than the model uncertainty.

As the distributions of the different uncertainty contributions are unknown, we assign them Gaussian distributions with $\sigma$-values of the uncertainties in Table~\ref{uncert}.
Since the various extraneous couplings are linear terms added to the measured force, and are expected to be uncorrelated, the uncertainty for each measured force is taken to be the quadrature sum of the relevant uncertainties.

The torque and the penultimate contributions are covariant with model uncertainty since we used the models to evaluate them all.
We do not account for these covariances as they are expected to be smaller than the corresponding variances and significantly smaller than the measurement uncertainties.

\begin{widetext}
\begin{center}		
	\begingroup
	\setlength{\tabcolsep}{10pt} 
	\renewcommand{\arraystretch}{1.5} 
	\begin{table}[ht!]
		\begin{center}
			\begin{tabular}{ |c|c|c|c|c| }
				\hline
				$1f$ (Hz) & $2f$ (Hz)& $F_{x}^{(2f)}$ ($pN$) & $3f$ (Hz) & $F_{x}^{(3f)}$ ($pN$) \\
				\hline
				4.16 & 8.32 & -- & 12.45 & 8.79$^{\pm0.32}_{\pm3.61\%}$\\
				5.17 & 10.34 & ~$16.59^{\pm2.33}_{\pm14.05\%}$ & 15.51 & $8.98^{\pm0.46}_{\pm5.07\%}$\\
				7.80 & 15.60 & $19.28^{\pm0.30}_{\pm1.56\%}$ & 23.40 & $9.12^{\pm0.37}_{\pm4.10\%}$\\
				8.55 & 17.11 & $19.10^{\pm0.47}_{\pm2.44\%}$ & 25.66 & $9.01^{\pm0.57}_{\pm6.36\%}$\\
				9.56 & 19.11 & $19.28^{\pm0.21}_{\pm1.12\%}$ & 28.67 & $9.15^{\pm0.24}_{\pm2.66\%}$\\
				\hline
				
			\end{tabular}
			\caption{Measured force amplitudes, $F_{x}^{(2f)}$ and $F_{x}^{(3f)}$, with corrections applied for extraneous couplings at the expected quadrupole $2f$ and hexapole $3f$ frequencies.}\label{RotationTable}
		\end{center}
	\end{table}
	\endgroup
\end{center}

\end{widetext}	

\begin{figure}[!h]
\includegraphics[width=0.45\textwidth]{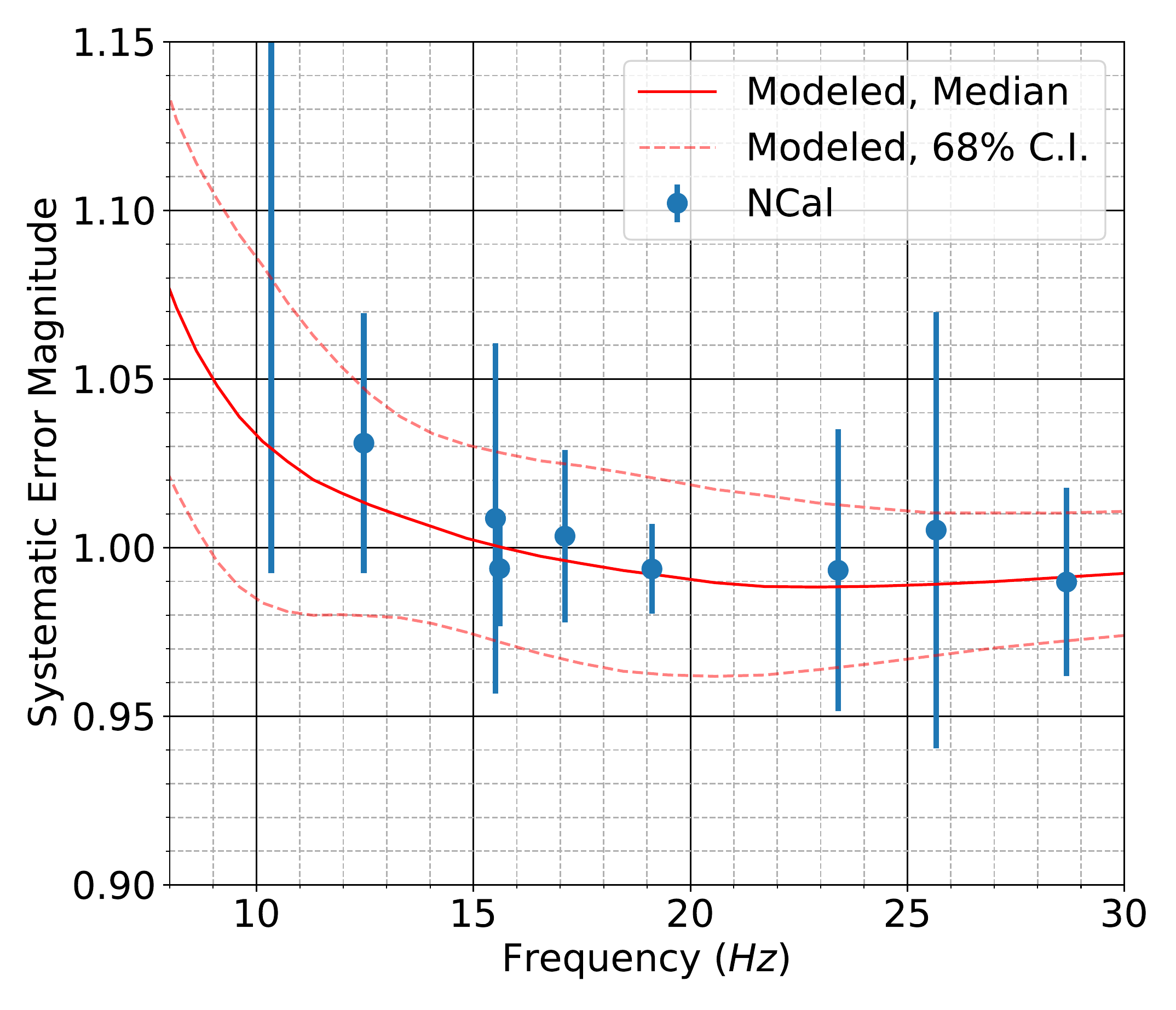}
\caption{Comparison between the modeled systematic error magnitude and systematic error as measured by the NCal.}\label{cal}
\end{figure}

\section{Absolute Calibration}

One immediate utility of these measurements is to act as a cross-check for the existing estimate of the systematic error in the observatory's strain readout. 
In \citet{O3Cal}, many reference measurements are concatenated to yield a model of the detector's strain response, $R_\text{model}$. The resulting model may be compared against the response measured with a direct force absolute reference, $R_\text{ref}$. The ratio of these two form a frequency dependent complex valued function that represents the systematic error in $R_\text{model}$:
\begin{equation}
\eta_{R} = \frac{R_{\text{ref}}}{R_{\text{model}}}. \label{eq:etaR}
\end{equation}

Ideally, if no systematic error is present then $\eta_{R}=1$. 
Figure \ref{cal} shows the systematic error as measured by the NCal system along with the modeled systematic error shown in  \citet{O3Cal}.
Here the NCal systematic error magnitude is determined  by taking the ratio of the modeled forces as predicted by the multipole model and the observed force corrected for systematics at each $2f$ and $3f$ frequency. Equation \ref{eq:dF_ASD} shows that the ratio of forces is the same as the ratio of strains.
Since we don't know the absolute phase of the NCal rotor, we are restricted to only evaluating the magnitude of the systematic error. 
The NCal results are consistent with the modeled systematic error at all frequencies.

By combining the modeled systematic error, the measured force, and the simulations of the geometry these measurements also allow a low-precision measurement of the gravitational constant detailed in Appendix \ref{BigG}.

%% file: 9-ConclusionAndFutureWork.tex
\section{Conclusion}
\label{Conclusion}

We have built and installed a gravitational calibration system at the LIGO Hanford detector which can reliably inject forces at known frequencies. 
These forces can be predicted to $<1\%$ relative uncertainty and have provided an independent check of the existing absolute calibration system.

In future work, we plan to analyze a set of simultaneous injections with both the NCal and the PCal systems. This will allow for two independent, absolute calibrations of the strain readout of the observatory which can be combined for increased precision. 

Although our current results are limited by the statistical uncertainty, we plan to decrease the various systematic effects with the goal of minimizing the total uncertainty in future NCal calibration injections. 
The influence of the torques on the force measurements can be decreased by altering the location of the NCal rotor and minimizing the interferometric beam offset on the test mass. 
We also plan to conduct dedicated environmental coupling measurements and mitigation to improve the magnetic and vibrational uncertainties.

The precision of the force model limits the NCal's absolute calibration independent of systematics. 
The current NCal prototype's model uncertainty is dominated by the precision of position surveying. 
With future upgrades (involving multiple rotors), we hope to circumvent the need for this surveying and thus decrease the uncertainty of the force model. 

Additionally, future iterations of this instrument may allow novel gravitational experiments such as searches for non-Newtonian gravity \citep{PhysRevD.84.082002}, terrestrial measurements of Shapiro delay \citep{Ballmer_2010, Sullivan_2020}, and measurements of the gravitational constant \citep{NCalGPaper}.
\\

%% file: 10-Acknowledgements.tex
\section{Acknowledgments}

This work was carried out at the Laser Interferometer Gravitational-Wave Observatory (LIGO) Hanford Observatory (LHO) by members of LIGO laboratory and the LIGO Scientific Collaboration including University of Washington, Seattle, University of Sheffield, and University of Glasgow. LIGO was constructed by the California Institute of Technology and Massachusetts Institute of Technology with funding from the National Science Foundation (NSF) and operates under Cooperative Agreement PHY‐0757058. Advanced LIGO was built under Award PHY‐0823459. Participation from the University of Washington, Seattle, was supported by funding from the NSF under Awards PHY-1607385, PHY-1607391, PHY-1912380, and PHY-1912514.

The authors would like to thank Peter Fritschel, Zsuzsanna Marka, Szabolcs Marka, and Dimitri Estevez for their helpful comments. We encourage readers to explore the rich theoretical and explorative history of gravitational force instruments in the citations, as decades of work by the cited groups have paved the way for the current LIGO NCal system's success.

%% file: A1-BigG.tex
\appendix
\section{Gravitational Constant Measurement}
\label{BigG}

In addition to the calibration results discussed here, the NCal system allows a unique measurement of the gravitational constant, $G$ \cite{NCalGPaper}. The pre-existing calibration system \cite{PCal} allows the force acting on the test mass to be measured precisely while the simulations described in Section \ref{sim} can account for the mass distributions. 

The observed forces at the different rotation frequencies were treated as independent measurements of $G$ once they were corrected for the known systematic error. These corrected measurements were then averaged to yield:
\begin{equation}
G=(6.56 \pm 0.41)\times10^{-11}\ \text{m}^3\ \text{kg}^{–1}\ \text{s}^{–2}
\end{equation}

Although not competitive with previous measurements \cite{bigG}, this is the first gravitational constant measurement utilizing a gravitational wave detector. 
We expect the precision of this measurement to increase with improvements to metrology of the geometry and to the accuracy of the force measurements.
However, it is unlikely that even future results with this technique will be competitive with other existing apparatus (torsion balances, etc.).